\documentclass[12pt]{article}
\usepackage{epsf}
\title{\bf QCD string in mesons and baryons}
\author{
D.S.Kuzmenko\thanks{e-mail: kuzmenko@heron.itep.ru}, Yu.A.Simonov
\thanks{e-mail: simonov@heron.itep.ru}}
\date{\it Institute of Theoretical and Experimental Physics,\\
117218, Moscow, Russia}
\newcommand{\be} {\begin{equation}}
\newcommand{\ee} {\end{equation}}
\newcommand{\bdm} {\begin{displaymath}}
\newcommand{\edm} {\end{displaymath}}
\newcommand{\bc} {\begin{center}}
\newcommand{\ec} {\end{center}}
\newcommand{\beqa} {\begin{eqnarray}}
\newcommand{\eeqa} {\end{eqnarray}}

\newcommand{\ver}{\mbox{\boldmath${\rm r}$}}
\newcommand{\veE}{\mbox{\boldmath${\rm E}$}}
\newcommand{\veB}{\mbox{\boldmath${\rm B}$}}
\newcommand{\ven}{\mbox{\boldmath${\rm n}$}}
\newcommand{\vex}{\mbox{\boldmath${\rm x}$}}

\begin{document}
\maketitle
\begin{abstract}
Field distributions generated by static $Q \bar Q$ and $QQQ$ sources are
calculated analyti- cally in the framework of Gaussian (bilocal) approximation of
Field Correlator Method. 
Special attention is paid to the $QQQ$ system and asymmetric configurations are
also studied. At large quark separations fields form distinct
stringlike shape. In both cases $Q \bar Q$ and $QQQ$ the string consists mainly 
of longitudinal color
electric field. Transverse color electric component contribution is shown to be
less then 3\%. Baryon string has an $Y$-like shape with a deep well at the 
string junction
position. Field distributions for quark-diquark and for three quarks on one line
 are considered. The interaction potential for quarks forming an
equilateral triangle is calculated. The material of the paper is illustrated by
24 3D colored pictures.
\end{abstract}

\section{Introduction}
 Field distributions inside the string connecting static $Q\bar Q$
sources have been measured repeatedly on the lattice using both
connected [1-3] and disconnected \cite{4,5} probes. Similar
measurements were done later also for Abelian projected
configurations \cite{6}. Analytic calculations for the disconnected
probe made in \cite{7} in the framework of the Gaussian
approximation to the Field Correlator Method (FCM)
 \cite{8,9} have revealed a clear
string-like structure of the same type as was found on the lattice.
The connected probe allows to measure the sign and modulus of each field
component separately.

A comparison of lattice data \cite{1} on magnitude and direction 
 of fields in the $Q \bar Q$ system with analytic predictions of 
 FCM was done in \cite{2}, demonstrating
 a remarkable agreement in all distributions. In particular, the
 measured decrease with distance of longitudinal electric field 
  from  the string axis ("the string profile") was
 remarkably well described by contribution of the lowest, (bilocal)
 correlator \cite{2}.  It should be noted that the input of FCM is
 the the field correlator, defined by scalar formfactors $D$ and 
 $D_1$ \cite{8}. The lattice measurements [10] yield for the latter 
 the  exponents 
 with the slope $T_g\approx 0.2 $fm. The dominance of the bilocal
 correlator (sometimes called the Gaussian Stochastic Model (GSM) of
 the QCD vacuum) was verified recently on the lattice by the precision
 measurement of static potential for $Q \bar Q$ Wilson loop in different SU(3)
 representations \cite{11}.  Analysis of data \cite{11} made in
 \cite{12} has demonstrated that GSM contribution around 99\% to the
 static  $Q\bar Q$ potential is consistent with the data.
 These results give an additional stimulus to the analytic
 calculation of field distributions using lowest bilocal correlator.

 In this paper we calculate $Q \bar Q$ and $QQQ$ field distributions in
 bilocal approximation of Field Correlator Method using connecting probes.
 We have found that the string consists mainly of longitudinal color electric
 field. For three quarks forming an equilateral triangle the string has an
 $Y$-like shape with a deep well at string junction position. We consider also 
 field
 distributions for quark-diquark and for three quarks at one line cases.
 Besides, we calculate three-quark-interaction potential for quarks forming an
 equilateral triangle. The well around the string-junction position provides
 a decrease of slope of the  interaction potential at small quark separations.
 
 Some results of this paper were presented earlier in [13].
 
 The paper contains 5 sections. In section 2 the $Q \bar Q$ longitudinal and 
 transverse field components distributions are derived analytically 
 and plotted in 3D pictures for different quark separations.
 Total field  distributions including perturbative one gluon exchange
  are also presented. In section 3 we derive analytically three quark field 
  distribution
 and plot it for different quark configurations. For three quarks forming an
 equilateral triangle we plot also total field including perturbative part.
  In section 4 baryon interaction potential is calculated. In section 5 the 
  results are
 summarized and their possible physical consequences are discussed.

\section{Field distributions in meson}
 
We shall study the field distributions using gauge invariant construction
$\rho_{\mu\nu} (x)$ consisting of probing plaquette $P_{\mu\nu}(x)$ and
Wilson loop $W$ connected with parallel transpoters $\Phi$ (see Fig.1). 
In what follows this construction is referred to as a connected probe.
By definition,

  \be 
    \rho_{\mu\nu} (x) = \frac{\langle  
    W^{\alpha}_{\beta}(x_0)\Phi^{\beta}_{\gamma}(x_0,x)
    (P_{\mu\nu}(x))^{\gamma}_{\delta}
    (\Phi^+)^{\delta}_{\alpha}(x,x_0)\rangle }{\langle W\rangle} -1,
 \label{1}
 \ee
 where
 \be
   W=\frac{1}{N_c}\mathrm{tr} P \exp(ig \oint_C A_{\mu}^at^adz_{\mu}),
 \label{2}
 \ee  
 \be
   (P_{\mu\nu}(x))^{\alpha}_{\beta}=
   (P \mathrm{exp}ig a^2 F_{\mu\nu}^a(x)t^a)^{\alpha}_{\beta},
 \label{3}
 \ee 
 \be
   \Phi^{\alpha}_{\beta} (x,y)=
   (P \mathrm{exp}ig \int_y^x A_{\mu}^at^adz_{\mu})^{\alpha}_{\beta},
 \label{4} 
 \ee
$W^{\alpha}_{\beta}(x_0)$ in (1) denotes an ordered exponential along contour 
$C$ of Wilson loop without the point $x_0=(0,0,0,0)$. A rectangular contour $C$
of size $R\times T$  lies in the plane (1~4)  (since quarks are static). 
The surface $S$ is bounded by the contour and has coordinates 
$x'=(x_1',x_2',x_3',x_4')$, where $0\leq x_1' \leq R,~  x_2' =x_3' \equiv 0, ~  
 -T/2\leq x_4' \leq T/2$.
The probing plaquette $P_{\mu\nu}(x)$ of size $a\times a$ is oriented in the 
plane ($\mu~\nu$). It is placed at a point $x=(x_1,x_2,x_3,x_4)$, where $x_1$ 
is the probe coordinate along
the string axis, $x_2$ is the distance from the probe to the string axis ,
$x_3=x_4\equiv 0$.  In the small-plaquette-size limit expanding
 (3) in powers of $a^2$ one obtains that the connected probe is propotional to $F(x)$:  
\be
 \rho_{\mu\nu} (x)  = iga^2 \frac{\langle  
    W^{\alpha}_{\beta}\Phi^{\beta}_{\gamma}(F_{\mu\nu}^a(x)t^a)^{\gamma}_{\delta}
    (\Phi^+)^{\delta}_{\alpha}\rangle }{\langle W\rangle}+O(a^4).
 \label{5}
\ee
Thus using the connected probe allows to measure $Q \bar Q$ color field components
not disturbed by probing plaquette in the small $a$ limit.
We shall calculate the connected probe (1) in bilocal approximation of FCM
 expanding Wilson loop in $F_{\mu\nu}$ and keeping only
bilocal correlators contribution.
Let us write Wilson loop in terms of $F_{\mu\nu}$ using the nonabelian Stokes
theorem:
 \be
   W=\frac{1}{N_c}\mathrm{exp}(ig \int_S F_{\mu\nu}(x,z_0) d\sigma_{\mu\nu}(x)),
 \label{6}
\ee  
where
\be
   F_{\mu\nu}(x,z_0)=\Phi(z_0,x) F_{\mu\nu}(x) \Phi(x,z_0).
\ee
Averaging Wilson loop over vacuum fields in bilocal approximation one has
\be
 \langle W \rangle
 \approx \mathrm{exp}(-\frac{g^2}{2}\int_S\int_S d\sigma_{\mu\nu}(x) d\sigma_{\rho\sigma}(x')
  \frac{1}{N_c}\mathrm{tr} \langle F_{\mu\nu}(x)\Phi(x,x') F_{\rho\sigma}(x') 
  \Phi(x',x)\rangle)
\label{8}
\ee
In bilocal approximation we arrived at a double surface integral.
The connected probe surface  $S_{\rho}$ consists of Wilson loop surface $S$ and
plaquette surface $S_P$,~ $S_{\rho}=S+S_P$, which gives for the  
double integral in (8) the sum of three terms,
\be
 \int_{S_{\rho}}\int_{S_{\rho}}=
 \int_S\int_S+2\int_S\int_{S_P}+ \int_{S_P}\int_{S_P}.
 \label{9}
\ee
Using (9) and (8) for the connected probe (1) we realize that the first term 
in (9)
 cancels with the denominator in (1) and third term of (9) disappears in 
the  small $a$ limit, yielding
$$
  \rho_{\mu\nu}(x)\simeq 
  \frac{\mathrm{exp}(-\frac{1}{2}\int_{S_{\rho}}\int_{S_{\rho}})}{\mathrm{exp}
  (-\frac{1}{2}\int_S\int_S)}-1=
$$
\be  
  =\mathrm{exp}(-\int_S\int_{S_P}-\frac{1}{2}\int_{S_P}\int_{S_P})-1=
  - \int_S\int_{S_P}+O(a^4),
\label{10}
\ee
where
$$
   -\int_S\int_{S_P}\equiv
   \rho_{\mu\nu}^{\mathrm{biloc}}(x)\equiv
$$
\be   
   \equiv -a^2 \int_S d\sigma_{\rho\sigma}(x')
  \frac{g^2}{N_c}\mathrm{tr} \langle F_{\rho\sigma}(x')\Phi(x',x) F_{\mu\nu}(x) 
  \Phi(x,x')\rangle.
\label{11}
\ee

Let us  define now an averaged (colorless) field strength at point $x$
in bilocal approximation as
$$ 
   \langle F_{\mu\nu}(x) \rangle_{Q \bar Q}\equiv 
   -\frac{1}{a^2}\rho_{\mu\nu}^{\mathrm{biloc}}(x)= 
$$
\be   
   =\int_S d\sigma_{\rho\sigma}(x')
  \frac{g^2}{N_c}\mathrm{tr} \langle F_{\rho\sigma}(x')\Phi(x',x) F_{\mu\nu}(x) 
  \Phi(x,x')\rangle.
\label{12}
\ee
 In FCM the following parametrization of bilocal correlators by scalar 
 formfactors $D$ and $D_1$ is suggested (see second ref. in \cite{8}):
$$
    \frac{g^2}{N_c}\mathrm{tr} \langle F_{\rho\sigma}(x')\Phi(x',x) F_{\mu\nu}(x) 
  \Phi(x,x')\rangle=  (\delta_{\rho\mu}\delta_{\sigma\nu}-
  \delta_{\rho\nu}\delta_{\sigma\mu})(D(h^2)+D_1(h^2))+
$$
\be
   +(h_{\mu}h_{\rho}\delta_{\nu\sigma}-h_{\mu}h_{\sigma}\delta_{\nu\rho}-  
    h_{\rho}h_{\nu}\delta_{\mu\sigma}+h_{\nu}h_{\sigma}\delta_{\mu\rho})
    \frac{\partial D_1(h^2)}{\partial h^2}\equiv D_{\rho\sigma,\mu\nu}(h),
\label{13}
\ee
where $h\equiv x-x'$.

In lattice measurements \cite{10} both formfactors were found to be exponential
beyond $x=0.2$fm with a slope $T_g\simeq0.2$fm:
\be
D(h^2) = D(0) \exp (-\mu|h|),
~~D_1(h^2) = D_1(0) \exp (-\mu|h|),
\label{14}
\ee
$$
   D_1(0) \simeq \frac13 D(0),~~ \mu\simeq 1 GeV ;~~ T_g\equiv \frac{1}{\mu}.
$$
We will use the form of $D$, $D_1$ (14) in the whole region of $x$ as it was
done in \cite{2}.

The values ($\rho~\sigma$)=(1~4) on the rhs of (12) are determined by the Wilson loop
orientation, and therefore the combinations of $D$ and $D_1$ entering
 (12) are:  
\be
  D_{14,i4}(h)=\delta_{1i}(D(h^2)+D_1(h^2))+(h_1h_i+h_4^2\delta_{1i})
  \frac{\partial D_1(h^2)}{\partial h^2},
\label{15}
\ee
\be
  D_{14,ik}(h)=(h_k h_4\delta_{i1}-h_ih_4\delta_{k1})
  \frac{\partial D_1(h^2)}{\partial h^2},
\label{16}
\ee
where $i,k=1,2,3$. One can see that (15) gives color electric field components
$\langle \veE(x)\rangle_{Q \bar Q}$, and (16) the color magnetic ones
 $\langle \veB(x)\rangle_{Q \bar Q}$.

Let us show that the only nonzero field components are
$\langle E_1(x)\rangle_{Q \bar Q}$ and $\langle E_2(x)\rangle_{Q \bar Q}$.
Really, in (15) $D_{14,34}(h)\equiv 0$ for $h_3\equiv x_3-x_3'\equiv 0$,
so that in (12) $\langle E_3(x)\rangle_{Q \bar Q}\equiv
\langle F_{34}(x)\rangle_{Q \bar Q}\equiv 0$. $D_{14,ik}(h)$ in (16) is 
antisymmetric in $h_4\equiv -x_4'$, so after integration
in (12) over $\int_S d\sigma_{\rho\sigma}(x')=\int_0^R dx_1'
\int_{-T/2}^{T/2}dx_4'$ one obtains $\langle \veB (x)\rangle_{Q \bar Q}
\equiv 0$.

Let us calculate $\langle E_1(x)\rangle_{Q \bar Q}$ and
$\langle E_2(x)\rangle_{Q \bar Q}$. 
\be
 \langle E_1(x)\rangle_{Q \bar Q}=\int_0^R dx_1'\int_{-T/2}^{T/2}dx_4'
 \left (D(h^2)+D_1(h^2)+(h_1^2+h_4^2)\frac{\partial D_1(h^2)}{\partial h^2}
 \right ) \equiv 
\label{17}
\ee
$$ 
 \langle E_1(x)\rangle_{Q \bar Q}^D+
 \langle E_1(x)\rangle_{Q \bar Q}^{D_1}.
$$
At $T\to \infty$
 
$$
\langle E_1(x) \rangle^D_{Q \bar Q} = 
D(0) \int_{x_1-R}^{x_1} dh_1 \int_{-\infty}^{\infty}dh_4 
\exp (-\mu  \sqrt{h_1^2+h_2^2+h_4^2})=
$$
\be
   2D(0) \int_{x_1-R}^{x_1} dx \sqrt{x^2+x_2^2}K_1(\mu\sqrt{x^2+x_2^2}),
\label{18}
\ee
where $K_1$ is McDonald function.
$$
   \langle E_1(x) \rangle^{D_1}_{Q \bar Q}= 
   \int_{x_1-R}^{x_1} dh_1 \int_{-\infty}^{\infty}dh_4
   \frac1 2 \left( \frac{\partial h_1 D_1}{\partial h_1}+
     \frac{\partial h_4 D_1}{\partial h_4} \right)=
$$
$$
   \frac{D(0)}{3}\int_0^{\infty}dh_4 
   \left( x_1 \exp(-\mu\sqrt{x_1^2+x_2^2+h_4^2})-
   (x_1-R)\exp(-\mu\sqrt{(x_1-R)^2+x_2^2+h_4^2})\right)=
$$
\be
    \frac{D(0)}{3} \left(x_1\sqrt{x_1^2+x_2^2}K_1(\mu\sqrt{x_1^2+x_2^2})-
    (x_1-R)\sqrt{(x_1-R)^2+x_2^2}K_1(\mu\sqrt{(x_1-R)^2+x_2^2})\right).
\label{19}
\ee

$$
\langle E_2(x)\rangle_{Q \bar Q}=\int_0^R dx_1'\int_{-T/2}^{T/2}dx_4'
h_1h_2\frac{\partial D_1(h^2)}{\partial h^2}=
  \int_{x_1-R}^{x_1}dh_1\int_{-\infty}^{\infty}dh_4
  \frac{h_2}{2}\frac{\partial D_1}{\partial h_1}=
$$
$$
    \frac{D(0)x_2}{3}\int_0^{\infty}dh_4 
   \left( \exp(-\mu\sqrt{x_1^2+x_2^2+h_4^2})-
   \exp(-\mu\sqrt{(x_1-R)^2+x_2^2+h_4^2})\right)=
$$
\be
    \frac{D(0)x_2}{3} \left(\sqrt{x_1^2+x_2^2}K_1(\mu\sqrt{x_1^2+x_2^2})-
    \sqrt{(x_1-R)^2+x_2^2}K_1(\mu\sqrt{(x_1-R)^2+x_2^2})\right)
\label{20}
\ee

Let us proceed with some evaluations. At $R\gg T_g$
\be
E_1^D(R/2,0)=4D(0)\int_0^{R/2}dx x K_1(x/T_g)= 2\pi D(0) T_g^2
\label{21}
\ee
At quark position
\be 
E_1^D(0,0)=2D(0)\int_0^{R}dx x K_1(x/T_g)= \pi D(0) T_g^2\equiv \sigma,
\label{22}
\ee
where $\sigma$=0.9GeV/fm is the $Q \bar Q$ string tension. 
The same value of $\sigma$ one has in the bilocal approximation of FCM from
area law of Wilson loop.

The fields $E_2$ and $E_1^{D_1}$ are maximal at $x\approx T_g$,
\be
E_2(0,T_g)=E_1^{D_1}(T_g,0)=\frac{D(0)}{3}T_g^2 K_1(1)=
\frac{0.38\pi}{6}D(0)T_g^2
\label{23}
\ee

We conclude that $E_2^{\max}/E_1^{\max}\approx 0.38/12=3.2\%$.

 Perturbative interaction in the leading $\alpha_s$ order is defined by the one
gluon exchange between quarks which leads to the Coulomb-type field contribution,
 \be
    \veE^{\mathrm{Coul}}=C_F\alpha_s
    \left(\frac{\ver_1}{r_1^3}-\frac{\ver_2}{r_2^3}\right),
\label{24}
\ee
where $\ver_1$ is a vector from the quark to the probe, and $\ver_2$ is from 
the antiquark to the probe. 
For fundamental quarks and $Q \bar Q$ in $SU(3)$ singlet the Casimir operator 
$C_F=\mathrm{tr}t^at^a=
\frac4 3$. The parameter $e=\frac4 3 \alpha_s=0.295$ is defined using the
lattice mesurements of $Q \bar Q$ potential fitted by Cornell potential
$V_{\mathrm{Corn}}=-\frac{e}{R}+\sigma R$ (see review [14] and references
therein).

In Figs.2--5(a)\footnote{You can get quality ps figures from
http://heron.itep.ru/$\sim$kuzmenko/figures.tar.gz}
 we plot $\langle E_1(x_1,x_2)\rangle_{Q \bar Q}^2$ distributions
for quark separations $R=T_g, 5T_g, 10T_g$ É $30T_g$. One can observe in the
pictures how the
 string of a characteristic shape between quark and antiquark is 
 forming starting from $R=5T_g$. On Figs. 2--5(b) the total field distributions
 with perturbative one
gluon exchange included are plotted for the same separations,
 \be
(\veE_{Q \bar Q}^{\mathrm{tot}})^2\equiv
 (E_1^{\mathrm{Coul}}+\langle E_1 \rangle_{Q\bar Q})^2 +
 (E_2^{\mathrm{Coul}})^2.
\label{25}
\ee
 One can see that the one gluon exchange dominates
at separations smaller then $T_g=0.2$fm.

 In Fig.6 $\langle E_2(x_1,x_2)\rangle_{Q \bar Q}^2$ distributions for
$R=T_g$ and $30T_g$ are presented. One can prove that $E_2$ does not create by
itself the string. The magnitude of $E_2$
at distances of the order of $T_g$ around the quark and antiquark positions 
is less then 3\% of $E_1^{\max}$ 
 and is decreasing fast with distance from $Q$ ($\bar Q$).

In Fig.7 are presented the transverse crossections of distributions  at fixed $x_1=R/2$ plotted in
Figs.2--5(a). For $R=10T_g$ and $30T_g$ the string profiles
practically coincide and at  $R\geq 10T_g$ the string acquires its saturation
form. The field magnitude in the middle of string is 
$E^{\mathrm{sat}}\equiv E_1^{\max}=1.8$GeV/fm$=2\sigma$.
The string width is $\delta x^{\mathrm{sat}}=2.2T_g$ (by definition of 
$\delta x^{\mathrm{sat}},
E_1^2(R/2,\delta x^{\mathrm{sat}}/2)=1/2(E^{\mathrm{sat}})^2$).

In Fig.8 a longitudinal string crossection at fixed $x_2=0$  for $R=30T_g$ 
is shown. The field is increasing rapidly in the range from $-3T_g$ to
 $3T_g$ and  then is forming the long plateau $E_1=E^{\mathrm{sat}}$.
In the same figure the total field with perturbative one gluon exchange included is
shown. It is decreasing monotonically at  $x_1>0$, flattens at
 $x_1\simeq 5T_g$ and is increasing at $25T_g<x_1<30T_g$;
   in the whole range of
  $0<x_1<R$ sign of derivative does not change.

In Fig.9 the field distributions $\langle E_1(x_1,x_2)\rangle_{Q \bar Q}^{D_1}$ 
 are presented for $R=T_g$ and $30T_g$. Just as
$E_2$, $E_1^{D_1}$ does also not create by itself the string.
Magnitude of $E_1^{D_1}$ at distances   $\sim T_g$ around quark and antiquark
positions  is 
less then 3\% $E^{\mathrm{sat}}$ and is decreasing fast
 off the quark (antiquark).
One can distinguish two symmetrical distributions around quark and antiquark.
 At $R=T_g$ they are superposed and the distribution has a maximum at zero.
At $R=30T_g$ a region in which the field is absent appears
 between them.
These distributions are antisymmetrical under transformations $x_1 \to -x_1$
and $(x_1-R)\to -(x_1-R)$ respectively.

\section{Field distributions in baryons}
In this section we study field distributions generated by three static quarks in
different configurations.
Let us first consider three quarks forming an equilateral triangle.
The contour of baryon Wilson loop $W^{(3Q)}$ consists of three contours
 $C_{\Gamma},~ \Gamma=A,B,C$,~  formed of quark trajectories,
 and the string junction trajectory (see Fig.10). The string junction position 
 is
 determined by the minimal area condition. In Fig.10 three contour
 lopes are intersecting at angles $\frac{2\pi}{3}$, forming a Mercedes star.
 The gauge invariance of the loop is provided by the antisymmetric tensors:
 \be
   W^{(3Q)}=\frac1 3 \epsilon_{\alpha\beta\gamma}\Phi^{\alpha}_{\alpha'}(C_A)
   \Phi^{\beta}_{\beta'}(C_B)\Phi^{\gamma}_{\gamma'}(C_C)
   \epsilon^{\alpha'\beta'\gamma'}
\label{26}
\ee   
 The probing plaquette  $P_{\mu\nu}(x)$ is attached by parallel transporters to
 the contour $C_A$ at a point $x^0: x_1^0=x_4^0=0,~x_3^0=R.$  
 For the entire connected probe construction the coordinate $~x_2\equiv 0$.

In the expression of averaged field in the bilocal FCM approximation (12) we are now
to sum over three surfaces $A,B,C:$
\be 
   \langle F_{\mu\nu}(x) \rangle_{3Q}
  = \sum_{\Gamma =A,B,C}\int_{\Sigma_{\Gamma}} d\sigma_{\rho\sigma}^{\Gamma}(x')
  \frac{g^2}{N_c}\mathrm{tr} \langle F_{\rho\sigma}(x')\Phi(x',x) F_{\mu\nu}(x) 
  \Phi(x,x')\rangle.
\label{27}
\ee
Here $\Sigma_{\Gamma}$ is the surface corresponding to the contour $C_{\Gamma}$,
$~\Gamma=A,B,C$; the sides of the rectangular surface have an extention $R$ and
$T$.
 Introducing in the plane $(x_1,x_3)$ unit vectors
 $\ven^A=(0,1), \ven^B=(\frac{\sqrt{3}}{2},-\frac1 2),
  \ven^C=(\frac{-\sqrt{3}}{2},-\frac1 2)$, we integrate in (27) over the surfaces:
\be
  d\sigma_{\rho\sigma}^{\Gamma}(x') F_{\rho\sigma}(x')=
  d\sigma_{i4}^{\Gamma}(x') F_{i4}(x')=
  n_i^{\Gamma}E_i(l'\ven^{\Gamma},x_4')dl'dx_4'
\label{28}
\ee
with the result 
\be
 \langle F_{\mu\nu}(x) \rangle_{3Q}=
 \sum_{\Gamma}n_i^{\Gamma}\int_0^Rdl'\int_{-T/2}^{T/2}dx_4'
 D_{i4,\mu\nu}(l'\ven^{\Gamma}-\vex,x_4'-x_4).
\label{29}
\ee

In what follows we  shall only consider the $D$ contribution which,
as  was demonstrated in previous section , gives the string form.
From (29),(15),(14) one obtains
$$
  \langle E_3(x_1,x_3) \rangle_{3Q}= D(0)\int_0^Rdl'\int_{-T/2}^{T/2}dx_4'
(e^{-\mu\sqrt{x_1^2+(x_3-l')^2+x_4'^2}}-
$$
\be
  -\frac1 2 e^{-\mu\sqrt{(x_1-\frac{\sqrt{3}}{2}l')^2+(x_3+l'/2)^2+x_4'^2}}-
  \frac1 2 e^{-\mu\sqrt{(x_1+\frac{\sqrt{3}}{2}l')^2+(x_3+l'/2)^2+x_4'^2}}),
\label{30}
\ee
$$
  \langle E_1(x_1,x_3) \rangle_{3Q}= D(0)\int_0^Rdl'\int_{-T/2}^{T/2}dx_4'  
  (\frac{\sqrt{3}}{2} e^{-\mu\sqrt{(x_1-\frac{\sqrt{3}}{2}l')^2+
  (x_3+l'/2)^2+x_4'^2}}-
$$
\be
 -\frac{\sqrt{3}}{2} e^{-\mu\sqrt{(x_1+\frac{\sqrt{3}}{2}l')^2+
  (x_3+l'/2)^2+x_4'^2}}).
\label{31}
\ee

As one can see from (30),(31), the baryon string is
a superposition of the meson string 
$\langle E_1(x_3,x_1) \rangle_{Q \bar Q}^D$ given in (18) and two such meson
strings obtained by rotation over angles $\frac{2\pi}{3}$ and $\frac{4\pi}{3}$:
\be
   \langle E_3(x_1,x_3) \rangle_{3Q}= \langle E_1(x_3,x_1) \rangle_{Q \bar Q}^D-
   \frac1 2  \langle E_1(x_3',x_1') \rangle_{Q \bar Q}^D-
   \frac1 2  \langle E_1(x_3'',x_1'') \rangle_{Q \bar Q}^D,
\label{32}
\ee
\be
   \langle E_1(x_1,x_3) \rangle_{3Q}=
\frac{\sqrt{3}}{2} \langle E_1(x_3',x_1') \rangle_{Q \bar Q}^D-
\frac{\sqrt{3}}{2}\langle E_1(x_3'',x_1'') \rangle_{Q \bar Q}^D,
\label{33}
\ee
where
\be
  x_1'=-\frac1 2 x_1-\frac{\sqrt{3}}{2}x_3,~~
  x_3'=\frac{\sqrt{3}}{2}x_1-\frac1 2 x_3;
\label{34}
\ee
$$
  x_1''=-\frac1 2 x_1+\frac{\sqrt{3}}{2}x_3,~~
  x_3''=-\frac{\sqrt{3}}{2}x_1-\frac1 2 x_3.
$$  

In Figs.11--13(a) and Fig.14 the squared field distribution is plotted
\be 
  \langle \veE(x_1,x_3) \rangle_{3Q}^2=
  \langle E_1(x_1,x_3) \rangle_{3Q}^2+
  \langle E_2(x_1,x_3) \rangle_{3Q}^2
\label{35}
\ee
for $R=T_g,5T_g,10T_g$ and $30T_g$. A characteristic  feature of
 all distributions
is a deep well around the string junction position. At the junction position itself
the field vanishes, since for the symmetry reason
at this point a preferred direction is absent. Outside the well the baryon
string becomes a sum of three meson strings going from the quarks to the string junction.
Let us note that field
distribution  in the region around string junction
will remain intact even for different distances from quarks to string junction 
were all the "outer" meson strings saturated.

The Coulomb field of one gluon exchange for three quarks is
\be
    \veE^{\mathrm{Coul}}_{3Q}=-\frac{C_F\alpha_s}{2} 
    \left( \frac{\ver_1}{r_1^3} + \frac{\ver_2}{r_2^3}+
    \frac{\ver_3}{r_3^3} \right),
\label{36}
\ee
where $\ver_i$ is distance from the $i$-th quark to the probe. Factor
$\frac{C_F}{2}$ appears due to contraction of antisymmetric tensors in Wilson loop
(26):
\be
 \frac{1}{3!}\epsilon_{\alpha\beta\gamma}\epsilon_{\alpha\sigma\tau}
 t_{\beta\sigma}^a t_{\gamma\tau}^a=-\frac1 6 \mathrm{tr}(t^at^a)=
 -\frac{C_F}{2},
\label{37}
\ee
where $C_F=\frac4 3$ is Casimir operator; $C_F\alpha_s=0.295$ (see explanation
 after (24)). The total field 
\be
   \veE^{\mathrm{tot}}_{3Q}= \langle \veE \rangle_{3Q}+
   \veE^{\mathrm{Coul}}_{3Q}
\label{38}
\ee   
is given in Figs.11--13(b) and Fig.15.
In Fig.11(b) the total field is at least two orders of magnitude greater than 
the nonperturbative
part (Fig.11(a)) in entire region considered. In Fig.12(b) we still see
a half of string, going out of a Coulomb spike. In Fig.13(b) and Fig.(15) the
Coulomb field only changes the shape of the adjacent piece of the string and 
does not disturb the rest of it. 

In Fig.16 are shown the longitudinal crossections of strings along $x_3$ axis at $x_1=0$ 
depicted in Figs.11--13(a), 14. 
At $R\geq 5T_g$ the shape of the well does not depend on quark separation.
  At $R\geq 10T_g$ the string becomes saturated; its longitudinal crossection
  at $0\leq x_3 \leq 6T_g$ rises from 0 to $(E^{\mathrm{sat}})^2$ 
 and then levels up. In the negative $x_3$ region 
 $0\leq |x_3| \leq 1.25T_g$ the crossection
 grows from 0 to 0.25GeV$^2$/fm$^2$ and 
 decreases rapidly at  $1.25T_g\leq |x_3| \leq 5T_g$.
 Let us define a well radius  $R_{\mathrm{well}}$ as
$E^2(0,R_{\mathrm{well}})=(E^{\mathrm{sat}})^2/2$. Then from the Fig.16 
we find $R_{\mathrm{well}}=1.75T_g$.
   
Let us consider now a quark-diquark configuration when two quarks are close to each
other and far from the third one. We take them to form an isosceles triangle
with a base much shorter than its lateral sides. The string is formed along the 
 minimal path, consistsing of three segments connecting quarks with string
junction and intersecting at angles $\frac{2\pi}{3}$, like a Mercedes star.
Let us denote $R_{QQ}$ the length of two short segments and $R_Q$ that of
the long one. In Eq.(29) to calculate the field distributions one can now integrate
over $dl'$ in the range from 0 to $R_Q$ for $\Gamma=A$ and from 0 to $R_{QQ}$ 
for $\Gamma=B,C$. Adding to  Eqs.(32),(33) the indices corresponding
to integration ranges, one has
$$
   \langle E_3(x_1,x_3) \rangle_{Q-QQ}= \langle E_1(x_3,x_1) 
   \rangle_{Q \bar Q}^{D,R_Q}-
   \frac1 2  \langle E_1(x_3',x_1') \rangle_{Q \bar Q}^{D,R_{QQ}}-
$$
\be   
   -\frac1 2  \langle E_1(x_3'',x_1'') \rangle_{Q \bar Q}^{D,R_{QQ}},
\label{39}
\ee
\be
   \langle E_1(x_1,x_3) \rangle_{Q-QQ}=
\frac{\sqrt{3}}{2} \langle E_1(x_3',x_1') \rangle_{Q \bar Q}^{D,R_{QQ}}-
\frac{\sqrt{3}}{2}\langle E_1(x_3'',x_1'') \rangle_{Q \bar Q}^{D,R_{QQ}}.
\label{40}
\ee
In Fig.17 we plot distributions
\be
\langle \veE(x_1,x_3) \rangle_{Q-QQ}^2=
\langle E_1(x_1,x_3) \rangle_{Q-QQ}^2+ \langle E_3(x_1,x_3) \rangle_{Q-QQ}^2
\label{41}
\ee
for $R_Q=30T_g$ and two values of $R_{QQ}$, (a): $R_{QQ}=0.5T_g$, (b):
$R_{QQ}=3T_g$. In  case (a) one observes a mesonic string  (cf. Fig.5(a)).
In case (b) one can see a baryonic string with the well in the string junction area.
How this transition happens? In accordance with (39),(40) the field strength at
 $x_1=x_3=0$ 
is determined by the difference of the field values  of mesonic strings of 
length $R_Q$ É $R_{QQ}$ at this point. The field strength of the mesonic string
(18) $E_0(R)$ at the origin for the string of length $R$ is
\be
   E_0(R)=\frac{2\sigma}{\pi}\int_0^{R/T_g}K_1(x)x dx.
\label{42}
\ee
This function increases linearly from zero at small $R$ and saturates to the 
asymptotic
value $E_{\mathrm{asymp}}=\sigma$ at $R>4T_g$. The dependence $E_0^2(R)$ 
is shown in Fig.18.
   
Let us define a radius of transition of baryonic string into mesonic one 
$R_{\mathrm{bar\to mes}}$ as 
 $E_0^2(R_{\mathrm{bar\to mes}})=E_{\mathrm{asymp}}^2/2$. From Fig.18 we find
 $R_{\mathrm{bar\to mes}}=1.5T_g$. At $R_{QQ}>R_{\mathrm{bar\to mes}}$ 
 the quark-diquark string has a characteristic baryonic form with the well at
 the string
 junction (the case of Fig.17(b)) and at $R_{QQ}<R_{\mathrm{bar\to mes}}$ 
 it turns into a mesonic one (the case of Fig.17(a)).

Let us now consider the case when quarks are placed along one line at distances
 $R_1$ and $R_2$ between them. The string junction position, as in previous
 cases, is defined by the minimal string length condition.
 For given quark positions the string length is minimal when the string junction
 position coincides with that of a middle quark (placed between two others)
  and is equal to 
 the sum of distances from this quark to others. 

The contour of Wilson loop is  in the plane  (1~4). Proceeding just as
in the beginning of section, we introduce in the plane (1~3) two unit
vectors $\ven_1=(-1,0), \ven_2=(1,0)$ and get
\be 
 \langle E_3(x_1,x_3) \rangle_{3Ql} \equiv 0,  
\label{43}
\ee
$$
 \langle E_1(x_1,x_3) \rangle_{3Ql}= -\int_0^{R_1}dl'\int_{-T/2}^{T/2}dx_4'  
 D_{14,14}(-l'-x_1,-x_3,x_4')+
$$
$$
+\int_0^{R_2}dl'\int_{-T/2}^{T/2}dx_4' D_{14,14}(l'-x_1,-x_3,x_4')=
$$
\be
=-\langle E_1(-x_1,x_3) \rangle_{Q \bar Q}^{D,R_1}+
\langle E_1(x_1,x_3) \rangle_{Q \bar Q}^{D,R_2}.
\label{44}
\ee
Distributions $\langle E_1(x_1,x_3) \rangle_{3Ql}^2$ for $R_1=R_2=15T_g$ and
$R_1=10T_g, R_2=20T_g$ are shown in Figs.19 (a),(b).
Since both mesonic strings  have saturated
 profiles, the field at the string junction position  $x_1=0$ is exactly zero.

\section{Quark interaction potential in baryon}

In this section we consider static quarks placed in vertices of an equilateral
triangle at distance $R$ from the string junction.

As the Wilson loop is a Green function of three static quarks,
it gives the quark interaction potential in the baryon,
\be
  V^{(3Q)}(R) =- \lim_{T\to \infty} \frac{1}{T}\ln \langle
  W^{(3Q)}(R,T)\rangle.
\label{45}
\ee 

Using bilocal approximation we get an expression of potential as a sum of 
surface integrals of bilocal correlators over Wilson loop surface. 
We integrate over the surfaces just as in calculation of averaged baryon field.
$$ 
   -\langle W^{(3Q)}(R,T) \rangle=   
  \frac12 \sum_{a,b=A,B,C}\int_{\Sigma_a}\int_{\Sigma_b}
  d\sigma_{\mu\nu}^a(x)d\sigma_{\rho\sigma}^b(x')\times
$$
$$  
 \times \frac{g^2}{N_c}\mathrm{tr} \langle F_{\mu\nu}(x)\Phi(x,x')F_{\rho\sigma}(x')\Phi(x',x) F_{\mu\nu}(x) 
  \rangle=
$$
\be  
=\frac32 \sum_{b=A,B,C}n_i^An_k^b\int_0^R\int_0^Rdldl'
\int_{-T/2}^{T/2}\int_{-T/2}^{T/2}dx_4dx_4'D_{i4,k4}(l\ven^A-l'\ven^b,x_4-x_4').
\label{46}
\ee

In the last equation of (46) we have used symmetry of the Wilson loop contour --
\be
   \sum_{a,b=A,B,C}=3\sum_{a=A;~ b=A,B,C}.
\label{47}
\ee

As in the previous section we shall only consider the contribution to the potential 
from the formfactor $D$. Then
$$     
V^{(3Q)}(R) =\lim_{T\to \infty} \frac3 2 \frac{D(0)}{T}\int_0^R\int_0^Rdldl'
\int_{-T/2}^{T/2}\int_{-T/2}^{T/2}dx_4dx_4'
\{e^{-\mu\sqrt{(l-l')^2+(x_4-x_4')^2}}-
$$
\be
 -e^{-\mu\sqrt{\frac3 4l'^2+(l+l'/2)^2+(x_4-x_4')^2}}\}.
 \label{48}
\ee 

Let us call the first exponential in (48) a "diagonal" as it is obtained at $a=b=A$ in (46)
and second one --- "nondiagonal" as it is obtained at $a=A,~ b=B,C$ in (46). 
Integrating in (48) one obtains
$$
  V^{(3Q)}(R) =\frac{6\sigma\mu^2}{\pi}\{R\int_0^Rdl lK_1(\mu l)-
  \frac{1}{\mu^3}(2-(\mu R)^2K_2(\mu R))-
$$
\be
  -\frac1 2 \int_0^Rdl \int_0^Rdl' \sqrt{\frac3 4l'^2+(l+l'/2)^2}
  K_1(\mu \sqrt{\frac3 4l'^2+(l+l'/2)^2})\},
\label{49}
\ee
where we have used the normalization of $D(0)$ (22).

First two terms of (49) result from the diagonal and the third one from
the nondiagonal exponentials of (48). The first term 
yields linear potential with a slope   $3\sigma$ at distances $R\gg T_g$, 
 the second is a small
correction to the first one (at large $R$ it equals $12\sigma T_g/\pi$).
The third term is increasing and significantly contributing to the potential
growth till  $R\simeq 3T_g$ and then flattens.
This term is due to  the well in the middle of baryonic string.

Potential behaviour $V^{(3Q)}(R)$ is shown in Fig.20(a). Tangent slope
$\sigma'(R)\equiv \frac{dV}{dR}(R)$ increases from zero to $3\sigma$ 
in the range of $0\leq R\leq 6T_g$. Beyond this range one can consider baryonic
string as a sum of three mesonic ones with the radius of well is subtracted
 (cf. Fig.8 and Fig.16).

It Fig.20(b) the total potential $V_{\mathrm{tot}}^{(3Q)}(R)$ is presented with
the one gluon exchange included,
\be
  V_{\mathrm{tot}}^{(3Q)}(R)=-\frac{C_F\alpha_s}{2}\sum \frac{1}{r_{ij}}+V^{(3Q)}(R)=
  -\frac{\sqrt{3}}{2}\, \frac{C_F\alpha_s}{R}+V^{(3Q)}(R).
\label{50}
\ee
$r_{ij}=R\sqrt{3}$ is quark separation, summation is performed over three
pairs, $C_F\alpha_s=0.295$ (see explanations after (24), (36)).
In the figure one can see a change of the potential slope as a consequence 
of the well in the middle of baryonic string.

Let us compare results of our calculations with  lattice measurements
of baryonic potential (review [14] and refs. therein). In the latter the 
potential
for quarks forming an equilateral triangle was fitted by a Coulomb plus
linear (Cornell) potential in the range of
$0.055$fm$\leq R\leq 0.71$fm. As a result the slope of the linear potential
 was found to be 2.6$\sigma$ which is equal to the $\sigma'(3.5T_g)$. Thus 
 the effect
of the potential decrease  due to the well in the middle of baryon string is 
consistent with  the effect  obtained on the lattice.

\section{Summary }

We have calculated connected-probe field distributions for
$Q\bar Q$ and $QQQ$ cases using the lowest (Gaussian) field
correlator. Since  the Gaussian correlator contributes 99\% to the static
$Q \bar Q$ potential [11,12], one expects that
higher correlators would not change our picture significantly.

The connected-probe analysis allows to distinguish directions of components of
color electric and color magnetic field due to probe
orientation. One can see that the string mainly consists of the longitudinal 
color electric field. The transverse electric component contributes
  less than 3\% of the longitudinal one. The magnitude of longitudinal component
 is given by the formfactor $D$ and the transverse one by $D_1$. The $D_1$
 contribution to longitudinal component is less than 3\%. At separations $R$
 greater
 then $10T_g$ the $Q \bar Q$ string profile becomes saturated and does not
 change with increase of $R$. The saturated string width is equal to $2.2T_g$.
  Our $Q \bar Q$ results are in
 agreement with earlier calculations in [1-3]. The bulk of the string fields is 
 also in accordance with disconnected-probe analyses \cite{4,5}.

The baryon string picture is obtained in our paper for the first time.
A deep well in the electric field distribution appears  around the 
string-junction position, where electric field vanishes  because of  symmetry 
arguments. The well has a radius of $1.75T_g$ demonstrating strongly suppressed fields 
in the middle of the heavy baryon. For quark-diquark configuration we have
defined the radius of baryonic to mesonic string transition, which was found
to be equal to $1.5T_g$. For
three quarks on one line  the field at the middle quark
position was found to vanish when  quark separations are large.
Since the Wilson  loop for the QQQ configuration used to generate
   interaction is also applicable for light quarks \cite{15}, physical
   consequences of this well can be in principle observable both for
   light and heavy hadrons. One consequence can be illustrated by Fig.20,
   where the nonperturbative  part of the potential grows very slowly
   at small $R$, so that the asymptotic slope is obtained only at
   very large distances. Therefore an effective  slope for
   ground-state hadrons can be some 10-20\% smaller, the fact which is in
agreement with relativistic quark    model of baryons \cite{16} and with
recent lattice calculations    of static QQQ potential \cite{14}.

One should note that the vanishing of fields at the string junction holds for
the connected-probe analysis of each separate field component, and may not be
true for field fluctuations at this point, measured by the disconnected probe.

Financial support of the grants 00-02-17836 and 00-15-96786 is gratefully
acknowledged.

\newpage
Figure 1: A connected probe for $Q \bar Q$.

Figure 2: $\langle E_1(x_1,x_2)\rangle_{Q \bar Q}^2$ (a) and
$(\veE_{Q \bar Q}^{\mathrm{tot}}(x_1,x_2))^2$ (b) distributions in
GeV$^2$/fm$^2$. Quark separation $R=T_g$. Both $x_1$ and $x_2$ are measured in
$T_g$. $Q$ and $\bar Q$ positions are marked
with  points.

Figure 3: $\langle E_1(x_1,x_2)\rangle_{Q \bar Q}^2$ (a) and
$(\veE_{Q \bar Q}^{\mathrm{tot}}(x_1,x_2))^2$ (b) distributions in
GeV$^2$/fm$^2$. $R=5T_g$. Both $x_1$ and $x_2$ are measured in
$T_g$. $Q$ and $\bar Q$ positions are marked with  points and vertical lines.

Figure 4: The same as in Fig.3 but for $R=10T_g$.

Figure 5: The same as in Fig.3 but for $R=30T_g$.

Figure 6: $\langle E_2(x_1,x_2)\rangle_{Q \bar Q}$ distribution in GeV/fm
for $R=T_g$ (a) and $R=30T_g$ (b). Both $x_1$ and $x_2$ are measured in
$T_g$. $Q$ and $\bar Q$ positions are marked with  points and vertical lines.

Figure 7: $Q \bar Q$ string transverse crossection profile 
$\langle  E_1(R/2,x_2) \rangle_{Q \bar Q}^2$ in GeV$^2$/fm$^2$. $x_2$ is
measured in $T_g$. Dotted, dashed, long dashed and solid lines refer to
$R=T_g,5T_g,10T_g$ and $30T_g$. At $R$ larger then $10T_g$ the string becomes 
saturated; $R=10T_g$ and $R=30T_g$ profiles practically coincide.

Figure 8: $\langle E_1(x_1,0)\rangle_{Q \bar Q}^2$ (solid line) 
and $(\veE_{Q \bar Q}^{\mathrm{tot}}(x_1,0))^2$ (dashed line) in GeV$^2$/fm$^2$.
$x_1$ is measured in $T_g$. $R=30T_g$.

Figure 9: $D_1$ contribution to $\langle E_1(x_1,x_2)\rangle_{Q \bar Q}$
distribution in GeV/fm for $R=T_g$ (a) and $R=30T_g$ (b). Both $x_1$ and $x_2$ 
are measured in $T_g$. $Q$ and $\bar Q$ positions are marked with  points and 
vertical lines.

Figure 10: A connected probe for $QQQ$.

Figure 11: $\langle \veE(x_1,x_3)\rangle_{3Q}^2$ (a) and
$(\veE_{3Q}^{\mathrm{tot}}(x_1,x_3))^2$ (b) distributions in GeV$^2$/fm$^2$ 
for quarks forming an equilateral triangle. Quark separation from the string
junction $R=T_g$. Both $x_1$ and $x_3$ are measured in $T_g$. Quark positions
 are marked with  points.

Figure 12: $\langle \veE(x_1,x_3)\rangle_{3Q}^2$ (a) and
$(\veE_{3Q}^{\mathrm{tot}}(x_1,x_3))^2$ (b) distributions in GeV$^2$/fm$^2$. 
 $R=5T_g$. Both $x_1$ and $x_3$ are measured in $T_g$. Quark positions
 are marked with  points and vertical lines.

Figure 13: The same as in Fig.12 but for $R=10T_g$.

Figure 14: The same as in Fig.12(a) but for $R=30T_g$.

Figure 15: The same as in Fig.12(b) but for $R=30T_g$.

Figure 16: $QQQ$ string profile $\langle E(0,x_2) \rangle_{3Q}^2$ in 
GeV$^2$/fm$^2$. $x_3$ is measured in $T_g$. Dotted, dashed, long dashed and
solid lines refer to $R=T_g,5T_g,10T_g$ and $30T_g$ respectively.

Figure 17: $\langle\veE(x_1,x_3)\rangle_{Q-QQ}^2$ distributions in 
GeV$^2$/fm$^2$ for quark-diquark configuration with $R_Q=30T_g, R_{QQ}=0.5T_g$
(a) and $R_Q=30T_g, R_{QQ}=3T_g$ (b). Both $x_1$ and $x_3$ are measured in 
$T_g$. Quark positions are marked with  points and vertical lines.

Figure 18: $E_0^2(R)\equiv (\langle E_1(0,0)\rangle_{Q \bar Q}^D)^2(R)$ 
distribution in GeV$^2$/fm$^2$ as a function of quark separation $R$ measured
in $T_g$.

Figure 19: $\langle E_1(x_1,x_3)\rangle_{3Ql}^2$ distribution in GeV$^2$/fm$^2$
for $R_1=R_2=15T_g$ (a) and $R_1=10T_g, R_2=20T_g$ (b). Both $x_1$ and $x_3$ 
are measured in $T_g$. Quark positions are marked with  points and vertical 
lines.

Figure 20: Three-quark-interaction potential $V^{(3Q)}(R)$ (a) and
the total potential with perturbative one-gluon-exchange included 
$V_{\mathrm{tot}}^{(3Q)}(R)$ (b) as functions of  distance $R$ from quarks to
string junction measured in $T_g$.


\begin{thebibliography}{99}


\bibitem{1}
A.Di Giacomo, M.Maggiore and S.Olejnik, Phys.Lett. {\bf B236}
(1990) 199; Nucl.Phys. {\bf B347} (1990) 441

\bibitem{2}
L.Del Debbio, A.Di Giacomo and Yu.A.Simonov, Phys.Lett. {\bf B332}
(1994) 111

\bibitem{3}
P.Cea and L.Cosmai, Phys.Rev. {\bf D52} (1995) 5152

\bibitem{4}
  G.S.Bali, Ch.Schlichter and K.Schilling, Nucl.Phys.Proc.Suppl. {\bf 42}
   (1995) 273; Phys.Rev. {\bf D51} (1995) 5165

 \bibitem{5}
 R.W.Haymaker, V.Singh, Y.Peng and J.Wosiek,
  Phys.Rev. {\bf D53} (1996) 389

  \bibitem{6}
  G.S.Bali, Ch.Schlichter and K.Schilling, Nucl.Phys.Proc.Suppl. {\bf
63} (1998) 519; Prog.Theor.Phys.Suppl. {\bf 131} (1998) 645

 \bibitem{7}
M.Rueter and H.G.Dosch, Z.Phys. {\bf C66} (1995) 245; \\
 H.G.Dosch, O.Nachtmann and M.Rueter, preprint HD-THEP-95-12, hep-ph/9503386 

  \bibitem {8}
  H.G.Dosch, Phys.Lett. {\bf B190}  (1987) 177;\\
  H.G.Dosch and Yu.A.Simonov, Phys.Lett. {\bf B205} (1988) 339;\\
   Yu.A.Simonov, Nucl.Phys.  {\bf B307}  (1988) 512

   \bibitem {9}
   Yu.A.Simonov,   Phys.Usp.   {\bf 39}  313 (1996)

   \bibitem{10}
   M.Campostrini, A.Di Giacomo and G.Mussardo, Z.Phys. {\bf C25}
   (1984) 173;\\
    A.Di Giacomo and H.Panagopoulos, Phys.Lett.  {\bf B285 } (1992)
    133;\\
    A.Di Giacomo, E.Meggiolaro and H.Panagopoulos, preprint IFUP-TH 12/96,\\
    hep-lat/9603017

   \bibitem{11}
   G.S.Bali, Nucl.Phys.Proc.Suppl. {\bf 83} (2000) 422, hep-lat/9908021

   \bibitem{12}
   Yu.A.Simonov, JETP Lett. {\bf 71} (2000) 187, hep-ph/0001244;\\
   V.I.Shevchenko and Yu.A.Simonov, hep-ph/0001299

    \bibitem{13}
   D.S.Kuzmenko and Yu.A.Simonov, Phys.Lett.{\bf B} (in press) hep-ph/0006192

   \bibitem{14}
   G.S.Bali, preprint HUB-EP-99-67, hep-ph/0001312

   \bibitem{15} Yu.A.Simonov, Phys.Lett. {\bf B228} (1989) 413;\\
   M.Fabre de la Ripelle and Yu.A.Simonov, Ann.Phys. (NY) {\bf 212}
   (1991) 235;\\
    B.O.Kerbikov and Yu.A.Simonov, Phys.Rev. D (in press) hep-ph/0001243

    \bibitem{16}
    S.Capstick and N.Isgur, Phys.Rev. {\bf D34} (1986) 2809
 

\end{thebibliography}
 \end{document}